\documentclass[journal]{IEEEtran}
\usepackage[utf8]{inputenc}
\usepackage{subfigure}
\usepackage{amssymb}
\usepackage{color,soul}
\usepackage[T1]{fontenc}
\usepackage{multicol}
\usepackage{multirow}
\usepackage{textcomp}
\usepackage{multirow}
\usepackage{bigdelim}
\usepackage{graphicx}
\usepackage{epstopdf}
\usepackage{amssymb}
\usepackage{amsmath}
\usepackage{psfrag}
\usepackage{amsthm}
\usepackage{subfigure}
\usepackage{algorithmic,algorithm}
\usepackage{mathrsfs}
\usepackage{framed}
\usepackage{color}
\usepackage{multirow,enumerate}
\usepackage{bigdelim}
\definecolor{shadecolor}{rgb}{0.92,0.92,0.92}
\usepackage{soul}
\usepackage{color}
\usepackage{cuted}
\usepackage{cite}
\theoremstyle{definition}

\newtheorem{definition}{Definition}

\newtheorem{remark}{Remark}

\makeatletter
\newcommand{\vast}{\bBigg@{3.2}}
\newcommand{\Vast}{\bBigg@{4.5}}

\makeatother

\graphicspath{{Figure/}}

\begin{document}

\title{Semi-Definite Relaxation Based ADMM for Cooperative Planning and Control of Connected Autonomous Vehicles}

\author{
	Xiaoxue Zhang,
	Zilong Cheng, 
	Jun Ma,	
	Sunan Huang,
	Frank L. Lewis,~\textit{Life Fellow, IEEE,}
	and Tong Heng Lee	
	\thanks{X. Zhang, Z. Cheng, and T. H. Lee are with the NUS Graduate School for Integrative Sciences and Engineering, National University of Singapore, Singapore 119077 (e-mail:xiaoxuezhang@u.nus.edu; zilongcheng@u.nus.edu; eleleeth@nus.edu.sg).}
	\thanks{J. Ma is with the Department of Mechanical Engineering, University of California, Berkeley, CA 94720 USA (e-mail: jun.ma@berkeley.edu).}
	\thanks{S. Huang is with the Temasek Laboratories, National University of Singapore, Singapore, 117411 (e-mail: tslhs@nus.edu.sg).}	
	\thanks{F. L. Lewis is with the Automation and Robotics Research Institute, University of Texas at Arlington, Arlington, TX 76118 USA (e-mail: lewis@uta.edu).}
	\thanks{This work has been submitted to the IEEE for possible publication.
		Copyright may be transferred without notice, after which this version may
		no longer be accessible.}
}

\maketitle

\begin{abstract}
	This paper investigates the cooperative planning and control problem for multiple connected autonomous vehicles (CAVs) in different scenarios. In the existing literature, most of the methods suffer from significant problems in computational efficiency. Besides, as the optimization problem is nonlinear and nonconvex, it typically poses great difficultly in determining the optimal solution. To address this issue, this work proposes a novel and completely parallel computation framework by leveraging the alternating direction method of multipliers (ADMM). The nonlinear and nonconvex optimization problem in the autonomous driving problem can be divided into two manageable subproblems; and the resulting subproblems can be solved by using effective optimization methods in a parallel framework. Here, the differential dynamic programming (DDP) algorithm is capable of addressing the nonlinearity of the system dynamics rather effectively; and the nonconvex coupling constraints with small dimensions can be approximated by invoking the notion of semi-definite relaxation (SDR), which can also be solved in a very short time. Due to the parallel computation and efficient relaxation of nonconvex constraints, our proposed approach effectively realizes real-time implementation and thus also extra assurance of driving safety is provided. In addition, two transportation scenarios for multiple CAVs are used to illustrate the effectiveness and efficiency of the proposed method.
\end{abstract}


\section{Introduction}
With the rapid development of information and communication technologies as well as the improvement of computational resources, connected automated vehicles (CAVs) have become one of the critical components in the context of intelligent transportation systems~\cite{rizk2018decision}. Generally, vehicles are equipped with advanced sensors that provide detailed information about the driving environment and onboard computing chip for efficient computation. Besides, the vehicle-to-vehicle (V2V) and vehicle-to-infrastructure (V2I) communication modules are equipped to enable the sharing of information with other participants through vehicular ad hoc networks~\cite{li2017dynamical}. All these technologies and equipment make substantial contributions to the development of intelligent transportation systems. However, cooperative planning and control (CPaC) in the tense traffic flow is still a long-standing challenge in the domain of autonomous driving~\cite{duan2020hierarchical}. Notably, the main task of the CPaC is to generate high-quality trajectories that satisfy the requirements resulted from the road geometry, collision avoidance, vehicle dynamics, and traffic rules during different driving tasks~\cite{xu2018cooperative}.

Researches of single-vehicle planning and control have been already studied extensively, and they can be referred to in numerous works~\cite{rasekhipour2016potential, chu2012local, ma2020data, chen2017constrained,amos2018differentiable,ma2020alternating,zhang2020trajectory}. On the other hand, the CPaC can be generally solved by learning-based and optimization-based approaches. Among all learning-based approaches, the reinforcement learning (RL) is quite effective to obtain the optimal or near-optimal action sequence through searching and evaluation~\cite{guan2020centralized,chu2019multi,ren2020improving}. However, it still suffers from certain shortcomings. For example, the number of the public dataset for autonomous driving is not enough and the data obtained from simulators cannot be generalized to all driving scenarios, and thus to obtain enough training dataset that contains all road condition and driving scenarios is still an open question so far. Besides, in some complex driving scenarios, determining a proper reward function, which is a critical component for the RL, is also not straightforward~\cite{bouton2019cooperation}. Some researches utilize the neural network to fit the real samples, which might ignore some corner cases in autonomous driving. This is because the value function in real traffic environment is closely related to the specific scenario, and not only the current state of all traffic participants need to be estimated, but also the states of all traffic participants in the future are required to be considered carefully~\cite{chen2019model}.

On the other hand, in terms of optimization-based approaches, the CPaC can be generally formulated as a constrained optimization problem that aims to generate collision-free trajectories. The trajectories should satisfy the constraints resulted from the vehicle dynamics, road geometry, collision avoidance, and traffic rules, meanwhile certain criterias such as comfort and safety should also be taken into account. There are already some optimization-based approaches available in the literature. For instance, \cite{burger2018cooperative} utilizes a mixed integer quadratic programming (MIQP) to solve the problem of cooperative trajectories planning for CAVs. \cite{mirheli2019consensus} proposes a distributed cooperative control to obtain collision-free trajectories for CAVs by formulating it as a mixed-integer nonlinear programming problem. In~\cite{lee2012development}, a nonlinear constrained optimization problem is formulated for an intersection scenario and solved by the active set method. However, most of these researches merely focus on simple motion tasks with simplified vehicle models. The approaches are no longer effective in more realistic and complex situations, where frequent interactions and coordination due to the vehicle model's strong nonlinearity and expansive computation are involved. Additionally, one major challenge in the cooperative optimization problem lies in how to handle the coupling constraints among connected vehicles with efficient computation. In general, this problem is nonlinear and nonconvex, which significantly increases the difficulty in deriving the optimal solution. As a result, the high complexity and nonconvexity of these constraints certainly induce heavy burdens to the computation efficiency~\cite{borrelli2006milp}.

Nowadays, the alternating direction method of multipliers (ADMM), which solves a convex optimization problem by breaking it into smaller and manageable ones, has become a considerable technique with remarkable scalability. It has been recently found the broad applicability in various areas, such as optimal control, distributed computation, machine learning, and so on~\cite{ma2020symmetric,zhou2019energy, zheng2016fast, ma2020arxiv}. Remarkably, the ADMM can solve a convex optimization problem with converging to a global optimum and achieve the parallel computation after decomposition, which exceedingly alleviates the typical computational burden resulted from the dimension growth of the optimization problem. Such advantages of the ADMM prompt many researchers to turn their attention to the nonconvex optimization problem. For example,~\cite{xu2016empirical} investigates the practical performance of the ADMM on several nonconvex applications and indicates that the ADMM also performs well on various nonconvex problems. \cite{hong2016convergence} analyzes the convergence of the ADMM when solving some specific nonconvex consensus and sharing problems. Moreover, an accelerated hierarchical ADMM algorithm is proposed in~\cite{zhang2020accelerated} for the nonconvex optimization problem. In~\cite{wang2019global}, a multi-block ADMM is presented to minimize a nonconvex and nonsmooth objective function subject to specific coupling linear equality constraints. In these past research works, the ADMM has been reasonably established at the theoretical level. Indeed, these advanced optimization techniques bring promising prospects to the area of autonomous driving.

This paper presents a novel ADMM-based approach for solving a nonconvex optimization problem with coupling constraints for CPaC of multiple CAVs. In this work, a nonlinear and nonconvex optimization problem is formulated, and a consensus ADMM is utilized to split the optimization problem into two small-scale subproblems, one with nonlinear dynamics constraints and the other one with nonconvex coupling constraints. The first subproblem considering the strong nonlinearity of the vehicle dynamics can be resolved by the differential dynamic programming (DDP) in a parallel manner. The second subproblem with the coupling nonconvex constraints for multiple CAVs is suitably addressed using several methods in parallel, including the semi-definite relaxation (SDR) and MIQP. Compared with some of the optimization approaches in the literature, this work can effectively relieve the computation burden arising from the nonlinearity and nonconvexity, which makes the real-time implementation possible and provides extra assurance of driving safety.

The remainder of this paper is organized as follows. Section~\ref{section:preliminary} gives the notation and the preliminary related to the DDP. Section~\ref{section:problem_formulation} defines the consensus nonlinear and nonconvex optimization problem with the introduction of the dynamic model, objective function, and constraints in the autonomous driving task. Section~\ref{section:optimization_in_parallel} proposes the ADMM algorithm for solving such a consensus optimization problem in a parallel framework. In Section~\ref{section:simulation}, two complex driving scenarios in autonomous driving are used to show the effectiveness of the proposed methodology. At last, the discussion and conclusion of this work are given in Section~\ref{section:discussion_and_conclusion}.

\section{Preliminary}
\label{section:preliminary}
\subsection{Notation}
The following notations are used in the remaining text. $\mathbb R^{a\times b}$ denotes the set of real matrices with $a$ rows and $b$ columns, $\mathbb R^{a}$ means the set of $a$-dimensional real column vectors. $A^{\top}$ and $x^{\top}$ denote the transpose of the matrix $A$ and vector $x$, respectively. $x \succ y$ and $x \succeq y$ denote that vector $x$ is element-wisely greater and no less than the vector $y$, respectively. $X\succ 0$ and $X\succeq 0$ represent that the matrix $X$ is positive definite and positive semi-definite, respectively.  $\mathbf 0_{a}$ and $\mathbf 0_{(a,b)}$ represent the $a$-dimensional all-zero vector and the $a$-by-$b$ all-zero matrix, respectively. $\mathbf 1_{a}$ and $\mathbf 1_{(a,b)}$ are the $a$-dimensional all-one vector and the $a$-by-$b$ all-one matrix, respectively. $I_{a}$ denotes the $a$-dimensional identity matrix. The Frobenius inner product is denoted as $\langle X, Y\rangle$, i.e., $\langle X, Y\rangle = \operatorname{Tr}(X^{\top}Y)$ for all $X,Y\in \mathbb R^{a\times b}$. The operator $\|X\|$ is the Euclidean norm of matrix $X$. The Kronecker product is denoted by $\otimes$. $\mathbb Z_{a}$ and $\mathbb Z_{a}^{b}$ represent the sets of positive integers $\{1,2,\cdots,a\}$ and $\{a,a+1,\cdots,b\}$ with $a<b$, respectively. $\operatorname{blockdiag}(X_1, X_2, \cdots, X_n)$ denotes a block diagonal matrix with diagonal entries $X_1, X_2, \cdots, X_n$; $\operatorname{diag}(a_1, a_2, \cdots, a_n)$ is a diagonal matrix with diagonal entries $a_1, a_2, \cdots, a_n$. $\left(\left\{x_i\right\}_{\forall i \in \mathbb Z_1^n}\right)$ denotes the concatenation of the vector $x_i$ for all $i\in \mathbb Z_1^n$, i.e., $\left(\left\{x_i\right\}_{\forall i \in \mathbb Z_1^n}\right) = \begin{bmatrix}
x_1^\top & x_2^\top & \cdots & x_n^\top \end{bmatrix}^\top = (x_1, x_2, \cdots, x_n)$.

\subsection{Differential Dynamic Programming}
The DDP is a second-order shooting method with a quadratic convergence rate, and it is typically deployed in trajectory optimization problems for nonlinear systems~\cite{tassa2012synthesis}. Notably, it calculates the second-order derivatives of the dynamic function and the cost function.

For a discrete-time dynamic model $x_{\tau+1} = f(x_\tau, u_\tau)$,  the cost function is defined as
\begin{IEEEeqnarray}{rCl}
	J_\tau(x,U_\tau) = \sum\limits_{i = \tau}^{T-1} \ell(x_i,u_i) + \ell_T(x_T),
\end{IEEEeqnarray}
where $U_\tau = \{u_\tau, u_{\tau+1}, \cdots, u_{T-1}\}$ denotes the sequence of control inputs from the time stamp $\tau$ to $T-1$, and $T$ is the prediction horizon. Define the value function at time $\tau$ as the optimal cost, where 
\begin{IEEEeqnarray}{rCl}
	V_\tau(x) = \min\limits_{U_\tau} \; J_\tau (x,U_\tau).
\end{IEEEeqnarray}
Also, it is pertinent to note that the value function of the terminal time stamp $T$ is $V_T(x) = \ell_T(x_T)$. According to the dynamic programming principle, the minimization problem over $U_\tau$ can be reduced into a sequence of minimization problems over one-step control, which is given by
\begin{IEEEeqnarray}{rCl}
	V_\tau(x) = \min\limits_{u_\tau} \; \ell(x_\tau,u_\tau) + V_{\tau+1}(f(x_\tau,u_\tau)).
\end{IEEEeqnarray}
Then, the perturbed Q-function is given by
\begin{IEEEeqnarray*}{rCl}
	\label{eq:perturbation}
	& & Q_{\tau}(\delta x_{\tau}, \delta u_{\tau}) \\
	&=& V_{\tau+1}(f(x_\tau+\delta x_\tau, u_\tau+\delta u_\tau)) - V_{\tau+1}(f(x_\tau, u_\tau)) \\
	& & + \ell_{\tau}(x_\tau+\delta x_\tau, u_\tau+\delta u_\tau)-\ell_{\tau}(x_\tau, u_\tau) \yesnumber
\end{IEEEeqnarray*}
where $\delta x_\tau$ and $\delta u_\tau$ denote the change of states and inputs at the time stamp $\tau$. Here, we expand~\eqref{eq:perturbation} into its second-order form as
\begin{IEEEeqnarray*}{rCl}
	\label{eq:perturbation_second_order}
	& & Q_{\tau}(\delta x_\tau, \delta u_\tau) \\
	&\approx& \frac{1}{2}\left[\begin{array}{c}
		1 \\
		\delta x_\tau \\
		\delta u_\tau
	\end{array}\right]^{\top}\left[\begin{array}{ccc}
		0 & \left(Q_{\tau}^{\top}\right)_{x} & \left(Q_{\tau}^{\top}\right)_{u} \\
		\left(Q_{\tau}\right)_{x} & \left(Q_{\tau}\right)_{x x} & \left(Q_{\tau}\right)_{x u} \\
		\left(Q_{\tau}\right)_{u} & \left(Q_{\tau}\right)_{u x} & \left(Q_{\tau}\right)_{u u}
	\end{array}\right]\left[\begin{array}{c}
		1 \\
		\delta x_\tau \\
		\delta u_\tau
	\end{array}\right] \IEEEeqnarraynumspace \yesnumber
\end{IEEEeqnarray*}
where
\begin{IEEEeqnarray*}{rCl}
	\label{eq:backward_1}
	\left(Q_{\tau}\right)_{x} &=&\left(\ell_{\tau}\right)_{x}+f_{x}^{\top}\left(V_{\tau+1}\right)_{x} \\
	\left(Q_{\tau}\right)_{u} &=&\left(\ell_{\tau}\right)_{u}+f_{u}^{\top}\left(V_{\tau+1}\right)_{u} \\
	\left(Q_{\tau}\right)_{x x} &=&\left(\ell_{\tau}\right)_{x x}+f_{x}^{\top}\left(V_{\tau+1}\right)_{x x} f_{x}+\left(V_{\tau+1}\right)_{x} \cdot f_{x x} \\
	\left(Q_{\tau}\right)_{u x} &=&\left(\ell_{\tau}\right)_{u x}+f_{u}^{\top}\left(V_{\tau+1}\right)_{x x} f_{x}+\left(V_{\tau+1}\right)_{x} \cdot f_{u x} \\
	\left(Q_{\tau}\right)_{u u} &=&\left(\ell_{\tau}\right)_{u u}+f_{u}^{\top}\left(V_{\tau+1}\right)_{x x} f_{u}+\left(V_{\tau+1}\right)_{x} \cdot f_{u u}. \IEEEeqnarraynumspace  \yesnumber
\end{IEEEeqnarray*}
Minimizing~\eqref{eq:perturbation_second_order} with respect to $\delta u_\tau$, we have
\begin{IEEEeqnarray*}{rCl}
	\label{eq:control_policy}
	\delta u_\tau^\star = \operatorname{argmin}_{\delta u_\tau} Q_{\tau}(\delta x_\tau, \delta u_\tau) = k_\tau + K_\tau \delta x_\tau,	
\end{IEEEeqnarray*}
where 
\begin{IEEEeqnarray*}{rCl}
	\label{eq:backward_2}
	k_\tau &=& -\left(Q_{\tau}\right)_{u u}^{-1}\left(Q_{\tau}\right)_{u} \\
	K_\tau &=& -\left(Q_{\tau}\right)_{u u}^{-1}\left(Q_{\tau}\right)_{u x}. \yesnumber
\end{IEEEeqnarray*}
By substituting this control policy~\eqref{eq:control_policy} into~\eqref{eq:perturbation_second_order}, we have 
\begin{IEEEeqnarray*}{rCl}
	\label{eq:backward_3}
	\Delta V_\tau &=& -\frac{1}{2} Q_{u,\tau} Q_{uu,\tau}^{-1} Q_{u,\tau} \\
	V_{x, \tau}   &=& Q_{x,\tau}-Q_{u,\tau} Q_{uu,\tau}^{-1} Q_{ux,\tau} \\
	V_{xx,\tau}   &=& Q_{xx,\tau}-Q_{xu,\tau} Q_{uu,\tau}^{-1} Q_{ux,\tau} \yesnumber
\end{IEEEeqnarray*}
By computing~\eqref{eq:backward_3} and the control policy terms $k_\tau,K_\tau $ gradually until $\tau = 0$, it constitutes the process named the backward pass. Subsequently, a forward pass is carried out to compute a new trajectory by 
\begin{IEEEeqnarray*}{rCl}
	\label{eq:forward}
	u_\tau &=& \hat{u}_\tau+\alpha k_\tau+K_\tau (x_\tau-\hat{x}_\tau) \\
	x_{\tau+1} &=& f(x_\tau, u_\tau),  \yesnumber
\end{IEEEeqnarray*}
where $\alpha$ is a backtracking search parameter. Normally, it is set to 1 and then reduced gradually.
Given an initial nominal trajectory $\{\hat{x}_\tau, \hat{u}_\tau\}$, the trajectory will be refined towards the optimal one after certain iterations of the backward pass and forward pass.

\section{Problem Formulation}
\label{section:problem_formulation}

\subsection{Network of Connected Vehicles}
In this paper, an undirected graph $\mathcal G(\mathcal V, \mathcal E)$ can be utilized to represent the constraint or information topology of multiple CAVs. The node set $\mathcal V=\{1,2, \cdots, N\}$ denotes the agents and $N$ is the number of agents. The edge set $\mathcal E = \{ 1,2, \cdots, M\}$ denotes the coupling constraints (information flow) between two interconnected vehicles, where $M$ denotes the number of agents and coupling constraints in the multi-agent system. The edge set $\mathcal E$ is defined as 
\begin{equation}
\begin{cases}
(i,j) \in \mathcal E(t), & \; d_{\mathrm{safe}} \leq \|p_i-p_j\|\leq d_{\mathrm{cmu}}\\
(i,j) \notin \mathcal E(t), & \; \text{otherwise},
\end{cases}
\end{equation}
where $d_{\text{cmu}}$ and $d_{\text{safe}}$ mean the maximum communication distance and minimum safe distance between two agents, respectively; $p_i$ and $p_j$ are the position vectors of the $i$th agent and $j$th agent. According to the communication topology, an adjacency matrix, which is denoted by $\mathcal D$, can be defined as a square symmetric matrix to represent the finite undirected graph $\mathcal G(\mathcal V, \mathcal E)$. The elements of $\mathcal D$ indicate whether pairs of vertices are adjacent/connected or not in the graph. Therefore, the neighbor nodes of the $i$th vehicle are the corresponding indexes of nonzero elements of the $i$th row in $\mathcal D$, which is represented by $\nu (i) = \{j|(i,j)\in \mathcal E, \forall j\in \mathcal V\}$.

\subsection{Problem Description}

\subsubsection{Dynamic Model}
In terms of an autonomous vehicle, we define the state vector as $x\in \mathbb R^n$ and the input vector as $u \in \mathbb R^m$. The general vehicle dynamic model for the $i$th vehicle can be represented by 
\begin{IEEEeqnarray}{rCl}
	x_{i(\tau+1)} = f(x_{i\tau}, u_{i\tau})
\end{IEEEeqnarray}
where $f$ denotes the kinematic dynamics of the $i$th vehicle, and $\tau$ is the time stamp. Note that the position vector $x_p\in \mathbb R^{n_p}$ is included in the state vector $x$, i.e., $x = (x_p, \cdots)$.

\subsubsection{Cost Function}
The cost function of the $i$th vehicle is 
\begin{IEEEeqnarray}{rCl}
	\sum_{\tau=0}^{T-1}  \left\|x_{i(\tau+1)} - x_{r,i(\tau+1)}\right\|_{Q_i}^2 + \left\|u_{i\tau}\right\|_{R_i}^2,
\end{IEEEeqnarray}
where $Q_i$ and $R_i$ is the weighting matrices and $x_{r,i}$ is the reference state vector that the vehicle needs to track. Here, the first term of the cost function penalizes the deviation between the state vector $x_i$ and the corresponding reference state vector $x_{r,i}$, and the second term penalizes the magnitude of the control input variable $u_i$. Note that the cost function is a convex, closed, and proper function.

\subsubsection{Constraints}
The restrictions on the state and input variables should be taken into consideration; hence the box constraint of input variable is introduced as
\begin{IEEEeqnarray}{rCl}
	\underline u_{i\tau} \preceq &u_{i\tau}& \preceq \overline u_{i\tau},
\end{IEEEeqnarray}
where $\underline u_{i\tau}, \overline u_{i\tau}$ denote the minimum value and maximum value of the input variables, respectively. Besides the box constraints, the collision avoidance constraints for the connected vehicles also need to be satisfied, which gives
\begin{IEEEeqnarray}{rCl}
	\left\| p_{i(\tau+1)} - p_{j(\tau+1)} \right\| \geq d_{\text{safe}}, \forall j\in \nu(i), \forall \tau \in \mathbb Z_0^{T-1}, 
\end{IEEEeqnarray}
where $p_i$ is the position vector for the $i$th vehicle, i.e., $p_{i\tau} = \begin{bmatrix} p_{x,i\tau} & p_{y,i\tau} \end{bmatrix}^\top$.

\subsubsection{Problem Formulation}
For all connected vehicles in the network, i.e., $\forall i\in \mathcal V$, each vehicle is required to satisfy the dynamic constraint and box constraints, as mentioned above. Besides, CAVs need to satisfy collision avoidance constraints. Hence, the multi-agent cooperative automation problem can be formulated as an optimal control problem, which is defined as
\begin{IEEEeqnarray*}{rCl}
	\label{eq:original_problem}
	\min &\quad&  \sum\limits_{i\in \mathcal V} \sum_{\tau=0}^{T-1}  \left\|x_{i(\tau+1)} - x_{r,i(\tau+1)}\right\|_{Q_i}^2 + \left\|u_{i\tau}\right\|_{R_i}^2 \\
	\operatorname{s.t.} && x_{i(\tau+1)} = f(x_{i\tau}, u_{i\tau}), \\
	&& \underline u_{i\tau} \preceq u_{i\tau} \preceq \overline u_{i\tau},\\
	&& \| p_{i(\tau+1)} - p_{j(\tau+1)} \| \geq d_{\text{safe}}, \\
	&& \forall \tau\in \mathbb Z_0^{T-1}, \forall j\in \nu(i), \forall i \in \mathcal{V}. \yesnumber
\end{IEEEeqnarray*}

Furthermore, we define the optimization variable $y\in \mathbb R^{NT(m+n)}$ as
\begin{IEEEeqnarray*}{rCl}
	y = 
	(\underbrace{y_{11}, \cdots, y_{1T}}_{y_1}, \cdots, \underbrace{y_{i1}, \cdots, y_{iT}}_{y_i}, \cdots, \underbrace{y_{N1}, \cdots, y_{NT}}_{y_N}),  
\end{IEEEeqnarray*}
where $y_i = \{y_{i1}, y_{i2}, \cdots, y_{i\tau}, \cdots, y_{iT}\} \in \mathbb R^{T(m+n)}$ for the $i$th vehicle, and $y_{i\tau} = ( x_{i\tau}, u_{i(\tau-1)} ) \in \mathbb R^{m+n}$. 

\begin{remark}
	In order to compute this problem in parallel, the host constraints and the coupling constraints in this collision avoidance multi-agent system are separated into two sets $\mathcal Y$ and $\mathcal Z$. The first set addresses the host constraints for each agent, and the other one aims to deal with the coupling constraints and box constraints.
\end{remark}

Define two sets $\mathcal Y$ and $\mathcal Z$ for the variable $y$ as
\begin{IEEEeqnarray*}{rCl}
	\label{eq:two_sets}
	\mathcal Y &=& \left\{
	y \in \mathbb R^{NT(m+n)} \middle| x_{i(\tau+1)} = f(x_{i\tau}, u_{i\tau})), \forall i\in \mathcal V
	\right\} \\
	\mathcal Z &=& \left\{
	z \in \mathbb R^{NT(m+n_p)} \middle|  
	\begin{array}{l}
		\underline u_{i\tau} \preceq z_{u,i\tau} \preceq \overline u_{i\tau}, \\
		\| z_{p,i\tau} - z_{p,j\tau} \| \geq d_\textup{safe} , \\
		\forall \tau \in \mathbb Z_{1}^T, \forall j\in \nu(i), \forall i \in \mathcal V		
	\end{array}
	\right\}, 
\end{IEEEeqnarray*}
where 
\begin{IEEEeqnarray*}{rCl}
	z = 
	\Big(\underbrace{z_{11}, \cdots, z_{1T}}_{z_1}, \cdots, \underbrace{z_{i1}, \cdots, z_{iT}}_{z_i}, \cdots, \underbrace{z_{N1}, \cdots, z_{NT}}_{z_N}\Big),  \IEEEeqnarraynumspace 
\end{IEEEeqnarray*}
with $z_{i\tau} = \left(z_{p,i\tau}, z_{u,i(\tau-1)}\right) \in \mathbb R^{m+n_p}$, $z_u = \left( \left\{z_{u,i\tau}\right\}_{\forall \tau \in \mathbb Z_{0}^{T-1}, \forall i \in \mathcal V} \right)$ which is the component of $z$, and the lower bound of $z_u$ is 
$\underline z_u = \left( \left\{\underline u_{i\tau}\right\}_{\forall \tau \in \mathbb Z_{1}^T, \forall i \in \mathcal V,} \right) \in \mathbb R^{NT(m+n)}$, and the upper bound of $z_u$ is $\overline z_u = \left( \left\{\overline u_{i\tau}\right\}_{\forall \tau \in \mathbb Z_{1}^T, \forall i \in \mathcal V,} \right)$. 
\begin{remark}
	  $x_{i(\tau+1)}$ and $u_{i(\tau)}$ are the components of the variable $y$ of state vector and the control input. $z_{p,i\tau}$ and $z_u$ are the components of the variable $z$.
\end{remark}

\begin{definition}
	For a non-empty set $\mathcal C$ and a variable $z$, the indicator function  is defined as $\delta_\mathcal C(z) = \begin{cases} 0, &\text{if } z\in \mathcal C \\ \infty, &\text{otherwise} \end{cases}$, where $\mathcal C$ is a convex set.
\end{definition}

Then, the optimal control problem~\eqref{eq:original_problem} can be converted into a more compact form as
\begin{IEEEeqnarray}{rCl}
	\label{eq:problem1}
	\min &\quad&  \sum\limits_{i\in \mathcal V} J(y) + \delta_{\mathcal Y}(y) + \delta_{\mathcal Z}(\mathcal T y),
\end{IEEEeqnarray}
where the matrix $\mathcal T = I_{NT} \otimes A\in \mathbb R^{NT(m+n_p) \times NT(m+n)}$ and the matrix $A = \begin{bmatrix} I_{n_p}  & \mathbf 0_{(n_p,n-n_p)} & \mathbf 0_{(n_p,m)} \\  \mathbf 0_{(m,n_p)} & \mathbf 0_{(m,n-n_p)} &  I_{m} \end{bmatrix}$.

By introducing a consensus variable $z\in \mathbb R^{NT(m+n_p)}$, we can rewrite \eqref{eq:problem1} as 
\begin{IEEEeqnarray}{rCl}
	\label{eq:problem2}
	\min &\quad& \sum\limits_{i\in \mathcal V} J_i(x_i,u_i) + \delta_{\mathcal Y}(y) + \delta_{\mathcal Z}(z) \nonumber \\
	\operatorname{s.t.} && \mathcal T y-z = 0 .
\end{IEEEeqnarray}

\section{Optimization in Parallel}
\label{section:optimization_in_parallel}
The augmented Lagrangian function of problem~\eqref{eq:problem2} is 
\begin{IEEEeqnarray*}{rCl}
	\label{eq:lagragian}
	&&   \mathcal L_\sigma(y,z;\lambda) \\
	&=&  J(y) + \delta_{\mathcal Y}(y) + \delta_{\mathcal Z}(z)  + \left\langle \lambda, E_py-z\right\rangle + \frac{\sigma}{2}\|\mathcal T y-z\|^2 \\
	&=&  J(y) + \delta_{\mathcal Y}(y) + \delta_{\mathcal Z}(z)  + \frac{\sigma}{2} \left\|\mathcal T y-z+\frac{\lambda}{\sigma}\right\|^2 - \frac{1}{2\sigma} \|\lambda\|^2  \\ \yesnumber 
\end{IEEEeqnarray*}
where $\lambda = (\lambda_1,\lambda_2,\cdots,\lambda_N)\in \mathbb R^{NT(m+n_p)}$ is the dual variable, and $\sigma$ is the penalty parameter.

It is straightforward to see that the ADMM algorithm can be denoted by
\begin{IEEEeqnarray*}{rCl}
	\label{eq:admm_iteration}
	y^{k+1} &=& \underset{y}{\operatorname{argmin}} \mathcal L_\sigma(y,z^k;\lambda^k) \IEEEyesnumber \IEEEyessubnumber \\
	z^{k+1} &=& \underset{r}{\operatorname{argmin}} \mathcal L_\sigma(y^{k+1},z;\lambda^k) \IEEEnonumber \IEEEyessubnumber \\
	\lambda^{k+1} &=& \lambda^k + \sigma(\mathcal T y^{k+1}-r^{k+1}), \label{subeq:update_lambda}\IEEEnonumber \IEEEyessubnumber 
\end{IEEEeqnarray*}
where the superscript $\cdot^k$ is the corresponding variable or parameter of the ADMM algorithm in the $k$th iteration. The stopping criterion is chosen in terms of the primal residual error, i.e., 
\begin{IEEEeqnarray}{rCl}
	\label{eq:stopping_criterion}
	\|\mathcal T y-z\|\leq \epsilon.
\end{IEEEeqnarray}

\subsection{Solving the First Sub-problem}
The first sub-problem of the ADMM algorithm is to determine the variable $y$ by
\begin{IEEEeqnarray*}{Rl}
	&\underset{y}{\operatorname{argmin}} \; \mathcal L_\sigma(y,r^k;\lambda^k) \\ =&\underset{y}{\operatorname{argmin}} \; J(y) + \delta_{\mathcal Y}(y) + \frac{\sigma}{2} \left\|\mathcal T y-z^k+\frac{\lambda^k}{\sigma}\right\|^2. \yesnumber
\end{IEEEeqnarray*}

Since there is no coupling term, this optimization sub-problem can be treated in a distributed manner for each agent $i\in \mathcal V$. 

For each agent $i$, we need to solve 
\begin{IEEEeqnarray*}{rCl}
	\label{eq:subproblem1}
	\min\limits_{y_i} &\quad& J_i(y_i) + \frac{\sigma}{2} \left\|\mathcal T_i y_i - z_i +\frac{\lambda_i}{\sigma}\right\|^2 \\
	\operatorname{s.t.} &\quad& x_{i(\tau+1)} = f(x_{i\tau}, u_{i\tau}),  \forall \tau\in \mathbb Z_0^{T-1}, \forall i \in \mathcal{V}, \yesnumber
\end{IEEEeqnarray*}
where the matrix $\mathcal T_i = I_{T} \otimes A \in \mathbb R^{T(m+n_p) \times T(m+n)}$, the variable $x_i = \left( \left\{x_{i(\tau+1)}\right\}_{\forall \tau\in \mathbb Z_{0}^{T-1}} \right)\in \mathbb R^{Tn}$ and $u_i = \left( \left\{u_{i\tau}\right\}_{\forall \tau\in \mathbb Z_{0}^{T-1}} \right) \in \mathbb R^{Tm}$. Since $y = (y_1, y_2,\cdots, y_i, \cdots, y_N)$, the variable $y_i\in \mathbb R^{T(m+n)}$ for the $i$th vehicle is
\begin{IEEEeqnarray*}{rCl}
	y_i = \Big\{ \underbrace{(x_{i1},u_{i0})}_{y_{i1}}, \cdots, \underbrace{(x_{i\tau},u_{i(\tau-1)})}_{y_{i\tau}}, \cdots, \underbrace{(x_{iT},u_{i(T-1)})}_{y_{iT}} \Big\}. 
\end{IEEEeqnarray*}
Based on the definition of the set $\mathcal Y$, given $x_{i0}$, the first sub-problem~\eqref{eq:subproblem1} can be rewritten as
\begin{IEEEeqnarray*}{rCl}
	\label{eq:subproblem1_1}
	\min &\quad&  \sum\limits_{i\in \mathcal V} \left\|x_{i} - x_{r,i}\right\|_{\hat Q_i}^2 + \left\|u_{i}\right\|_{\hat R_i}^2 + \delta_{\mathcal X_i}(x_{i}) + \delta_{\mathcal U_i}(u_i), \\
	\operatorname{s.t.} && x_{i(\tau+1)} = f(x_{i\tau}, u_{i\tau}),  \forall \tau\in \mathbb Z_0^{T-1}, \forall i \in \mathcal{V}, \yesnumber
\end{IEEEeqnarray*}
where the weighting matrices $\hat Q_i = I_T \otimes Q_i$ and $\hat R_i = I_T \otimes R_i$, the vector $x_{r,i}\in \mathbb R^{Tn}$ is the reference state vector, i.e., $x_{r,i} = \left( \left\{x_{r,i(\tau+1)}\right\}_{\forall \tau \in \mathbb Z_{0}^{T-1}} \right)$, $\delta_{\mathcal X_i}(\cdot)$ and $\delta_{\mathcal U_i}(\cdot)$ denote the indicator function with respect to the non-empty set $\mathcal X_i$ and $\mathcal U_i$, respectively, $\mathcal X_i =\left\{ x_{i} \mid \underline x_{i} \preceq x_{i} \preceq \overline x_{i} \right\}$ and $\mathcal U_i = \left\{ u_i \mid \underline u_{i} \preceq u_{i} \preceq \overline u_{i} \right\}$,  $\underline x_{i} = \left( \{\underline x_{i\tau}\}_{\forall \tau\in \mathbb Z_{0}^{T-1}} \right)$,  $\underline u_{i} = \left( \{\underline u_{i\tau}\}_{\forall \tau\in \mathbb Z_{0}^{T-1}} \right)$. Furthermore, the definitions of $\overline x_{i}$ and $\overline u_i$ are in the similar way as those of $\underline x_{i}$ and $\underline u_i$.

Hence, the problem~\eqref{eq:subproblem1_1} is the standard format such that the DDP algorithm can be adopted directly. The pseudocode of the DDP algorithm is shown in Algorithm~\ref{alg:DDP}.
\begin{algorithm}
	\caption{DDP Algorithm for the $i$th Vehicle}
	\label{alg:DDP}
	\begin{algorithmic}
		\STATE {\textbf{Initialization:} initial nominal trajectory $\left\{ \tilde x_{i\tau}, \tilde u_{i\tau}\right\}_{\forall \tau \in \mathbb Z_0^T}$; derivatives of the cost-to-go function $\ell_i$ and dynamic model $f_i$ for the $i$th vehicle; the maximum iteration number of DDP, i.e., $r_{\text{ddp}}$.}
		\STATE {Set the initial iteration step $r=0$.}
		\WHILE {$r\leq r_{\text{ddp}}$ \textbf{or} not meet stopping criterion}
		\STATE{}\COMMENT{$\triangleright$ \textit{Backward pass.}}
		\FOR {$\tau = T-1, \cdots, 0$}
		\STATE {compute~\eqref{eq:backward_1},~\eqref{eq:backward_2},~\eqref{eq:backward_3}}
		\ENDFOR\\
		\STATE {Set the backtracking line-search parameter $\alpha =1$. }\\
		\COMMENT {$\triangleright$ \textit{Forward pass.}}
		\STATE {Use~\eqref{eq:forward} to compute a new nominal trajectory.} 
		\STATE {Decrease $\alpha$.}
		\STATE {$r = r +1$.}
		\ENDWHILE
	\end{algorithmic}
\end{algorithm}

\subsection{Solving the Second Sub-problem}
The second sub-problem is 
\begin{IEEEeqnarray*}{rCl}
	\label{eq:sub_problem2}
	& \underset{z}{\operatorname{min}} \; \mathcal L_\sigma(y^{k+1},z;\lambda^k)= \min\limits_{z} \; \delta_{\mathcal Z}(z) + \frac{\sigma}{2} \left\|\mathcal T y^{k+1}-z+\frac{\lambda^k}{\sigma}\right\|^2. \\ \yesnumber
\end{IEEEeqnarray*}
Here, the second sub-problem~\eqref{eq:sub_problem2} is equivalent to
\begin{IEEEeqnarray*}{rCl}
	\min\limits_{z}     &\quad& \left\|\mathcal T y^{k+1} - z-\frac{\lambda^k}{\sigma}\right\|^2  \\
	\operatorname{s.t.} &\quad& \underline z_u \preceq z_u \preceq \overline z_u \\
	&& \| z_{p,i\tau} - z_{p,j\tau} \| \geq d_{\text{safe}}\\
	&& \; \forall \tau \in \mathbb Z_{1}^T, \forall j\in \nu(i), \forall i\in \mathcal V, \yesnumber
\end{IEEEeqnarray*}
where $z_u$, the lower bound and upper bound of the optimization variable $\underline z_u$ and $\overline z_u$ have been defined in~\eqref{eq:two_sets}. Besides, $z_{p,i\tau}$ is one of the component of $z_{i\tau}$ regarding the position vector, i.e., $z_{i\tau} = \left(z_{p,i\tau}, z_{u,i\tau}\right)$. 

On the other hand, the variable $z$ can be rewritten as $z = \left(z_1, z_2, \cdots, z_N\right) \in \mathbb R^{NT(m+n_p)}$ with $z_i=\left(z_{i1},z_{i2},\cdots, z_{i\tau},\cdots,z_{iT}\right)\in \mathbb R^{T(m+n_p)}$ and $z_{i\tau} = \left(z_{p,i\tau}, z_{u,i\tau}\right)$. Thus, the variable $z$ can be separated regarding the subscript $\cdot_\tau$, i.e., $z_\tau = \left( \left\{ z_{i\tau} \right\}_{\forall i\in \mathcal V} \right) \in \mathbb R^{N(m+n)}$. For each $\tau$, $z_{i,\tau} = (z_{p,i\tau}, z_{u,i\tau})$. Therefore, we can derive $z_{p,\tau} = \left( \left\{z_{p,i\tau}\right\}_{\forall i\in \mathcal V} \right) \in \mathbb R^{Nn_p}$, and $z_{u,\tau} = \left( \left\{z_{u,i\tau}\right\}_{\forall i\in \mathcal V} \right)\in \mathbb R^{Nm}$.

For all $\tau\in \mathbb Z_0^{T-1}$, We can separate this problem into $T$ problems. At the time stamp $\tau$, the problem is defined as
\begin{IEEEeqnarray*}{rCl}
	\label{eq:sub_problem2_2}
	\min\limits_{z_\tau} &\quad& \left\| z_{\tau} - \mathcal T_\tau y_{\tau}^{k+1} - \frac{\lambda_{\tau}^k}{\sigma} \right\|^2 +\delta_{\mathcal Z_{u,\tau}} (z_{u,\tau}) \\
	\operatorname{s.t.} && \| z_{p,i\tau}-z_{p,j\tau} \| \geq d_{\text{safe}} , \forall i\in \mathcal V, \forall j\in \nu(i), \yesnumber
\end{IEEEeqnarray*}
where $y_{\tau} = \left( \left\{y_{i\tau}\right\}_{\forall i\in \mathcal V} \right)\in \mathbb R^{N(m+n)}$ and $\lambda_{\tau} = \left( \left\{\lambda_{i\tau}\right\}_{\forall i\in \mathcal V} \right) \in \mathbb R^{m+n_p}$, and the set $\mathcal Z_{u,\tau} = \left\{z_{u,\tau} \in \mathbb R^{NTm} \mid \underline z_{u,\tau} \preceq z_{u,\tau} \preceq \overline z_{u,\tau} \right\}$. Since the constraints only consider the position component $z_{p,\tau}$ of the variable $z_\tau$, we can divide the problem~\eqref{eq:sub_problem2_2} into two separate parts according to the two components $z_{p,\tau}$ and $z_{u,\tau}$ of the variable $z_\tau$. 

The first part in terms of the variable $z_{u,\tau}$ is 
\begin{IEEEeqnarray}{rCl}
	\label{eq:sub_problem2_3}
	\min\limits_{z_{u,\tau}} &\quad& \left\| z_{u,\tau} - \mathcal T_{u,\tau} y_{\tau}^{k+1} - \frac{\lambda_{u,\tau}^k}{\sigma} \right\|^2 +\delta_{\mathcal Z_{u,\tau}} (z_{u,\tau}) 
\end{IEEEeqnarray}
where the matrix $\mathcal T_{u,\tau} = I_{N} \otimes \begin{bmatrix} \mathbf 0_{(m,n)} & I_{m} \end{bmatrix}$ is used to extract $y_{u,\tau}$ from $y_{\tau}$.  
It is straightforward to obtain
\begin{IEEEeqnarray}{rCl}
	\label{eq:solve_z_u}
	z_{u,\tau} = \operatorname{Proj}_{\mathcal Z_{u,\tau}} (\mathcal T_{u,\tau} y_{\tau}^{k+1} + \frac{\lambda_{u,\tau}^k}{\sigma}),
\end{IEEEeqnarray}
where the operator $\operatorname{Proj}_C(z)$ denotes the projection of the variable $z$ onto the set $C$. Here, the solution of~\eqref{eq:solve_z_u} can be easily solved by confining all elements of the vector $\mathcal T_{u,\tau} y_{\tau}^{k+1} + \frac{\lambda_{u,\tau}^k}{\sigma}$ to be inside the set $\mathcal Z_{u,\tau}$. If some elements of the vector $\mathcal T_{u,\tau} y_{\tau}^{k+1} + \frac{\lambda_{u,\tau}^k}{\sigma}$ are outside $\mathcal Z_{u,\tau}$, these elements are bounded by $\mathcal Z_{u,\tau}$. If some elements of the vector $\mathcal T_{u,\tau} y_{\tau}^{k+1} + \frac{\lambda_{u,\tau}^k}{\sigma}$ are inside $\mathcal Z_{u,\tau}$, these elements will maintain their value.

The second part for the avariable $z_{p,\tau}$ is
\begin{IEEEeqnarray*}{rCl}
	\label{eq:sub_problem2_4}
	\min\limits_{z_{p,\tau}} &\quad& \left\| z_{p,\tau} - \mathcal T_{p,\tau} y_{\tau}^{k+1} - \frac{\lambda_{p,\tau}^k}{\sigma} \right\|^2 \\
	\operatorname{s.t.} && \| z_{p,i\tau}-z_{p,j\tau} \| \geq d_{\text{safe}} , \forall i\in \mathcal V, \forall j\in \nu(i), \yesnumber
\end{IEEEeqnarray*}
where the matrix $\mathcal T_{p,\tau} = I_{N} \otimes \begin{bmatrix} I_{n_p} & \mathbf 0_{(n_p,m+n-n_p)} \end{bmatrix}$ is used to extract $y_{p,\tau}$ from $y_{\tau}$. 

Since the inequality constraints are nonconvex, the SDR can be used to solve the problem~\eqref{eq:sub_problem2_2}.
Though it is much easier to solve a relaxed optimization problem, the results of the relaxed problem determine the bounds of the optimal value of the original nonconvex problem. Therefore, the SDR is used to solve the nonconvex problem~\eqref{eq:sub_problem2_4}.  
Actually, this problem~\eqref{eq:sub_problem2_4} can be formulated as a quadratically constrained quadratic programming (QCQP) problem, which can be defined as
\begin{IEEEeqnarray*}{rCl}
	\label{eq:subproblem2_SDR1}
	\min\limits_{z_{p,\tau}} &\quad& \left(z_{p,\tau} - c_{p,\tau}\right)^\top \left(z_{p,\tau} - c_{p,\tau}\right)  \\
	\operatorname{s.t.} && (z_{p,i\tau} - z_{p,j\tau})^\top (z_{p,i\tau} - z_{p,j\tau}) \geq d_{\text{safe}}^2 \\
	&& \forall i\in \mathcal V, \forall j\in \nu(i), \yesnumber
\end{IEEEeqnarray*}
where $c_{p,\tau} = \mathcal T_{p,\tau} y_{\tau}^{k+1} + \frac{\lambda_{p,\tau}^k}{\sigma}$. By defining a matrix $M_{ij}\in \mathbb R^{n_p\times Nn_p}$ such that $M_{ij} z_{p,\tau} = z_{p,i\tau} - z_{p,j\tau}$, we can rewrite~\eqref{eq:subproblem2_SDR1} as 
\begin{IEEEeqnarray*}{rCl}
	\label{eq:subproblem2_SDR2}
	\min\limits_{z_{p,\tau}} &\quad& z_{p,\tau}^\top z_{p,\tau} -2c_{p,\tau}^\top z_{p,\tau}  \\
	\operatorname{s.t.} && z_{p,\tau}^\top K_{ij} z_{p,\tau} \geq d_{\text{safe}}^2 , \forall i\in \mathcal V, \forall j\in \nu(i), \yesnumber
\end{IEEEeqnarray*}
where $K_{ij} = M_{ij}^\top M_{ij} \in \mathbb S^{Nn_p}$. 

Note that two formulas in deriving an SDR problem are given by
	\begin{IEEEeqnarray*}{rCl}
		z_{p,\tau}^\top z_{p,\tau} &=& \operatorname{Tr}(z_{p,\tau} z_{p,\tau}^\top) \\
		z_{p,\tau}^\top K_{ij} z_{p,\tau} &=& \operatorname{Tr}(K_{ij} z_{p,\tau} z_{p,\tau}^\top ). \yesnumber
	\end{IEEEeqnarray*}
Thus, by introducing a new variable $Z_\tau = z_{p,\tau} z_{p,\tau}^\top \in \mathbb S^{Nn_p}$, the QCQP problem~\eqref{eq:subproblem2_SDR2} can be rewritten as
\begin{IEEEeqnarray*}{rCl}
	\label{eq:subproblem2_SDR3}
	\min\limits_{z_{p,\tau}} &\quad& \operatorname{Tr}(Z_\tau) -2c_{p,\tau}^\top z_{p,\tau}  \\
	\operatorname{s.t.} && \operatorname{Tr}(K_{ij}Z_\tau ) \geq d_{\text{safe}}^2 , \forall i\in \mathcal V, \forall j\in \nu(i) \\
	&&  Z_\tau = z_{p,\tau} z_{p,\tau}^\top. \yesnumber
\end{IEEEeqnarray*}
Here, the quadratic terms in~\eqref{eq:subproblem2_SDR2} have been converted into linear ones. Besides, a nonlinear equality constraint is introduced in~\eqref{eq:subproblem2_SDR3}. Then, the problem~\eqref{eq:subproblem2_SDR3} can be relaxed to a convex one by changing the last nonconvex equality constraint $Z_\tau = z_{p,\tau} z_{p,\tau}^\top$ for a convex inequality constraint $Z_\tau - z_{p,\tau} z_{p,\tau}^\top \succeq 0$. Notice that $Z_\tau - z_{p,\tau} z_{p,\tau}^\top$ can be formulated as a Schur complement, which results in a symmetric matrix given by $\begin{pmatrix} Z_\tau & z_{p,\tau} \\ z_{p,\tau}^\top & 1 \end{pmatrix}$. 
\begin{remark}
	By using the Schur complement, the positive semi-definite constraint becomes a second-order cone constraint which is easy to be handled.
\end{remark}

Thus, the problem can be obtained as
\begin{IEEEeqnarray*}{cCl}
	\label{eq:SDR}
	\min\limits_{z_{p,\tau},Z_\tau \in \mathbb S^{Nn_p}} &\quad& \operatorname{Tr}(Z_\tau) -2c_{p,\tau}^\top z_{p,\tau}  \\
	\operatorname{s.t.} &&  \operatorname{Tr}(K_{ij} Z_\tau) \geq d_{\text{safe}}^2, \forall i\in \mathcal V, \forall j\in \nu(i) \\
	&& \begin{pmatrix} Z_\tau & z_{p,\tau} \\ z_{p,\tau}^\top & 1 \end{pmatrix} \succeq 0, \yesnumber
\end{IEEEeqnarray*}
which is a semi-definite programming (SDP) problem, because one of the constraints is changed into a looser one. It is apparent that the optimal value of~\eqref{eq:SDR} is not greater than the optimal value of~\eqref{eq:subproblem2_SDR1}, because the cost function is minimized under a larger domain in~\eqref{eq:SDR}. Besides, if $Z_\tau = z_\tau z_\tau^\top$ at the optimum of the problem~\eqref{eq:SDR}, then $z_\tau$ will be optimal. Note that the feasible results of the relaxed problem represent that the collision avoidance constraints are satisfied. 
\begin{remark}
	It indicates that the optimality and the feasibility of a nonconvex QCQP problem may not be guaranteed by the SDR. Otherwise, an NP-hard problem would have been solved in a polynomial time, which is impossible based on the current state of the science. However, the result is still a non-trivial solution of a nonconvex QCQP problem.
\end{remark}

\begin{remark}
		Many practical experiences have already indicated that the SDR can provide accurate or near-optimal approximations. In some cases, the solution of the SDP is not a feasible solution to the nonconvex problem. In that case, we can employ randomization, which uses the optimal solution of the SDP to extract a feasible solution to the nonconvex problem~\cite{park2017general}. 
\end{remark}

\subsubsection{Comparison with Mixed Integer Quadratic Programming}
Here, the MIQP can be used as a method to solve the nonconvex problem~\eqref{eq:sub_problem2_4} for comparative purposes.
Assume the collision region is a polyhedron, which can be expressed as the intersection of $s$ half planes $\left\{x\in \mathbb R^{n_p} \mid Px\leq q \right\}$  where $P = \begin{bmatrix} P_1^\top & P_2^\top & \cdots & P_s^\top \end{bmatrix}^\top \in \mathbb R^{s\times n_p}$ with $P_i \in \mathbb R^{n_p}$ for $i\in \mathbb Z_{1}^s$ , and $q = \begin{bmatrix} q_1 & q_2 & \cdots & q_s \end{bmatrix}^\top \in \mathbb R^{s}$. Each row $P_i^\top x \leq q_i$ of $Px\leq q$ denotes one of the half plane which is to form the polyhedron. In order to avoid the collision, the big-M method is used to relax this problem. Here, in order to avoid this polyhedron (collision region), $x$ must be at least in one of the half planes, i.e., $P_1^\top x \leq q_1$ or $P_2^\top x \leq q_2$, or $\cdots$, or $P_s^\top x \leq q_s$. Since the union of logic operator ``or'' is hard to compute, it should be transformed into logic ``and'' operator, which is a convex form in an optimization problem. Thus, binary variables are introduced as
\begin{IEEEeqnarray*}{rCl}
	P_1^\top x &\geq& q_1 - M e_1\\
	P_2^\top x &\geq& q_2 - M e_2\\
	&\vdots& \\
	P_s^\top x &\geq& q_s - M e_s\\
	\sum_{i=1}^{s} e_i &\leq& s-1, \yesnumber
\end{IEEEeqnarray*}
where $e_i\in \{0,1\}$ is a binary variable and $M$ is a sufficiently large positive number. If $e_i=0$, the corresponding constraints are satisfied and if $e_i=1$, it is relaxed. The last constraint $\sum\limits_{i=1}^s e_i \leq s-1$ is used to guarantee that at least one constraint is satisfied. For example, if $s = 4$ or $8$, an obstacle can be represented as a rectangle or octagon by using $s$ binary variables. In the MIQP, the computational time largely depends on the number of integer or binary variables, i.e., $s$. Thus, the number of binary variables should be as small as possible to decrease the computational time. Here, we set $s=4$. We further define matrices $P_1, P_3$, which are used to extract the position variable  $x,y$ in three dimensions, and also matrices $P_2, P_4$, which are are used to extract the position variable with negative sign $-x,-y$ in three dimensions. $q_i = d_{\text{safe}}, \forall i\in \mathbb Z_1^s$. $e_i$ is the integer binary variables. Therefore, we can rewrite the constraints $\| z_{p,i\tau}-z_{p,j\tau} \| \geq d_{\text{safe}}$ as 
\begin{IEEEeqnarray}{rCl}
	P_{ijr} z_{p,\tau} \geq d_{\text{safe}} - Me_{ijr}
\end{IEEEeqnarray}
by introducing binary variables $e_{ijr}$ for each original $g_{ij}$ inequality constraint. Thus, the subproblem can be transformed to
\begin{IEEEeqnarray*}{rCl}
	\label{eq:MIQP}
	\min\limits_{z_{p,\tau}} &\quad& \left\| z_{p,\tau} - \mathcal T_{p,\tau} y_{\tau}^{k+1} - \frac{\lambda_{p,\tau}^k}{\sigma}\right\|^2 + \delta_{\mathcal C}(e)  \\
	\operatorname{s.t.} && P_{ijr} z_{p,\tau} \geq d_{\text{safe}} - Me_{ijr} \\
	&& \sum\limits_{r=1}^6 r_{ijr} \geq 5\\
	&& \forall i\in \mathcal V, \forall j\in \nu(i), \yesnumber
\end{IEEEeqnarray*}
where $e = \left( \left\{e_{ijr}\right\}_{\forall r\in \mathbb Z_1^s, \forall j\in \nu(i),\forall i\in \mathcal V } \right) \in \mathbb R^{s\sum_i(r_i)}$, $P_{ijr}$ is used to extract the position variable of $z_{i\tau}-z_{j\tau}$ in one dimension with positive or negative sign, the set $\mathcal C$ is a nonconvex cone with two integer elements 0,1, i.e., $\mathcal C = \{0,1\}^{s\sum_i r_i}$. 

\subsubsection{Comparison with Interior Point Method}
Another comparative study is performed by solving the nonconvex problem~\eqref{eq:sub_problem2_4} using the interior point OPTimizer (IPOPT) method, which is a very comprehensive approach to solve the nonlinear nonconvex programming problem.

\subsection{Proposed Algorithm}
The pseudocode of our proposed algorithm is shown in Algorithm~\ref{alg:ADMM}. Note that all 
\begin{algorithm}
	\caption{ADMM for Multi-Vehicle Cooperative Automation}
	\label{alg:ADMM}
	\begin{algorithmic}
		\STATE {\textbf{Initialization:} dynamic model for all agents $i\in \mathcal V$; communication network $\mathcal G(\mathcal V, \mathcal E)$;  parameters $M$, $P_{ij\tau}$; weighting matrices $Q_i$ and $R_i$  in the cost function; upper and lower bound of the state $x$ and control input $u$, i.e., $\overline x, \underline x$ and $\overline u, \underline u$, respectively; the maximum iteration number of ADMM, i.e., $k_{\text{admm}}$; initial state $x_{i,0}$ at initial time $\tau = 0$ for all vehicle $i\in \mathcal V$; penalty parameter $\sigma>0$; variables $y^0\in \mathbb R^{NT(m+n)}$, $z^0 \in \mathbb R^{NT(m+n_p)}$; dual variable $\lambda^0 \in \mathbb R^{NT(m+n_p)}$. }
		\STATE {Set the outer iteration $k = 0$.}
		\WHILE {$k\leq k_{\text{admm}}$ \textbf{or} stopping criterion~\eqref{eq:stopping_criterion} is violated}
		\STATE {Update $y^{k+1}$ in parallel for $i\in \mathcal V$ via the DDP algorithm.}
		\STATE {Update $z_{u}^{k+1}$ in parallel for $\tau \in \mathbb Z_0^T$ by~\eqref{eq:solve_z_u}}.
		\IF {Use the SDR}
		\STATE {Update $z_{p}^{k+1}$ in parallel for $\tau \in \mathbb Z_0^T$ by solving the problem~\eqref{eq:SDR}.}
		\ELSIF {Use the MIQP}
		\STATE {Update $z_{p}^{k+1}$ in parallel for $\tau \in \mathbb Z_0^T$ by solving the problem~\eqref{eq:MIQP}.}
		\ELSIF {Use the IPOPT}
		\STATE {Update $z_{p}^{k+1}$ in parallel for $\tau \in \mathbb Z_0^T$ by solving the problem~\eqref{eq:sub_problem2_4} directly.}
		\ENDIF
		\STATE {Update $\lambda^{k+1}$ by~\eqref{subeq:update_lambda}.}
		\STATE {$k=k+1$.}
		\ENDWHILE
	\end{algorithmic}
\end{algorithm}

\section{Simulation Results}
\label{section:simulation}
\subsection{Dynamic Model of the Vehicle}
The dynamic model of the $i$th vehicle can be characterized by 
\begin{IEEEeqnarray}{rCl}
	p_{x,i(\tau+1)} &=& p_{x,i\tau}+f_r \left(v_{i\tau}, \delta_{i\tau}) \cos (\theta_i(\tau)\right) \label{subeq:px} \IEEEyesnumber \IEEEyessubnumber \\
	p_{y,i(\tau+1}) &=& p_{y,i\tau}+f_r \left(v_{i\tau}, \delta_{i\tau}) \sin (\theta_i(\tau)\right) \label{subeq:py} \IEEEnonumber \IEEEyessubnumber \\
	\theta_{i(\tau+1)} &=& \theta_{i\tau}+\sin^{-1} \left(\frac{\tau_s v_{i\tau} \sin (\delta_{i\tau})}{b_i}  \right) \label{subeq:theta} \IEEEnonumber \IEEEyessubnumber \\
	v_{i\tau+1} &=& v_{i\tau}+ \tau_s a_{i\tau} \IEEEnonumber \IEEEyessubnumber 
\end{IEEEeqnarray}
where the subscript $\cdot_i$ means the corresponding parameters or variables for the $i$th vehicle, the subscript $\cdot_\tau$ means the corresponding parameters or variables at the time stamp $\tau$, $p_x$ and $p_y$ are the position of the center point of the vehicle in $X$ and $Y$ dimension in the Cartesian coordinates, respectively, $\theta$ represents the heading angle of the vehicle with 0 in the positive X-dimension, $v$ denotes the velocity of the vehicle, $\delta$ is the steering angle, and $a$ is the acceleration of the vehicle, $b$ denotes the wheelbase of this vehicle, $\tau_s$ is the sampling time, and the function $f_r(v,\delta)$ is 
\begin{IEEEeqnarray*}{rCl}
	f_r(v, \delta) &=& b+\tau_sv \cos (\delta)-\sqrt{b^{2}- \left(\tau_s v\sin(\delta)\right)^{2} }.
\end{IEEEeqnarray*}

Define the state vector as $x= \begin{bmatrix} p_x & p_y & \theta & v \end{bmatrix}^\top$ and the input vector as $u = \begin{bmatrix}\delta & a \end{bmatrix}^\top$. The vehicle dynamic model can be rewritten as 
\begin{IEEEeqnarray}{rCl}
	x_{i(\tau+1)} = f(x_{i\tau}, u_{i\tau}).
\end{IEEEeqnarray}

\subsection{Simulation Results}
Here, we focus on CPaC for multiple connected vehicles under the scenario of intersections in autonomous driving. In this paper, two scenarios are considered, including a three-way junction scenario and an intersection scenario. The optimization algorithm is implemented in a PC with Intel(R) Xeon(R) CPU E5-1650 v4 @ 3.60GHz, and all the programs are conducted in Python 3.7.

In the simulation, the steering angle $\delta$ is confined into the range $[-0.6, 0.6]$ rad, and the acceleration is within $[-3,3]$ m/s$^2$. The length and width of all vehicles are 2.5 m and 1.6 m, respectively. The penalty parameter $\sigma$ of the augmented Lagrangian function is set as 10. Also, the sampling time $\tau_s$ is 0.1 s. The maximum iteration numbers of the DDP algorithm and the ADMM algorithm are set as 100 and 100, respectively. The safety distance $d_{\text{safe}}$ is defined as $3$ m. The lane width of all roads is set to 4 m. The initial trajectory of the DDP algorithm is the initial states with control inputs being 0. The prediction horizon $T$ is set to be 100. The tolerance of stopping criterion in~\eqref{eq:stopping_criterion} is set to be 0.01. 

\subsubsection{Scenario 1 (Three-way Junction)}
\begin{figure}[t]
	\centering
	\includegraphics[width=1\linewidth]{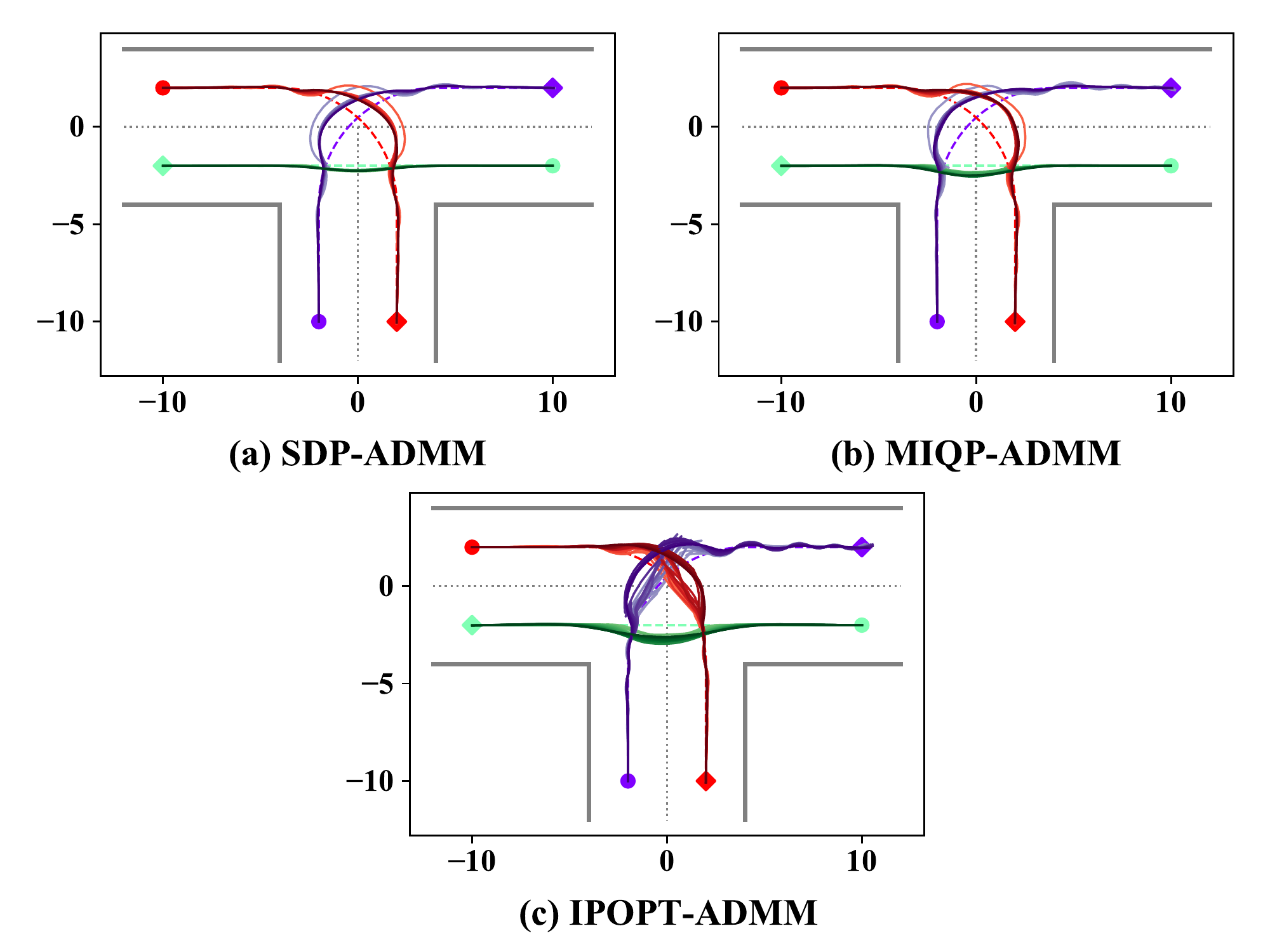}
	\caption{Trajectories of the 3 vehicles in all ADMM iterations when using different methods to solve the second subproblem.}
	\label{fig:trajs_all_iters_in_all_solvers}
\end{figure}

In the scenario of a three-way junction, three vehicles are used to show the performance in cooperative planning and control. Here, the three vehicles have three types of behaviors (turning left, turning right, and going straight) in three lanes. All subfigures in Fig.~\ref{fig:trajs_all_iters_in_all_solvers} illustrate the trajectories of the three vehicles in all ADMM iterations, with different solvers when solving the second subproblem. In this figure, the thick solid black lines are the road boundaries, and the thin dotted gray lines are the road center-lines to separate lanes in different directions. The circle marker and diamond marker denote the start point and end point of a vehicle. The reference trajectories of the three vehicles are represented by dashed lines. All trajectories for one vehicle in all ADMM iterations are represented by a group of similar colors. For example, the first group of purple line denotes the group of one vehicle's trajectories in all ADMM iterations. Also, the darker the color, the more iterations it represents. In this figure, we can observe that the trajectories become smoother with the increase of iterations. Here, the SDP-ADMM (our proposed method), MIQP-ADMM (comparison 1), and IPOPT-ADMM (comparison 2) are the ADMM approach with the use of the SDP, MIQP and IPOPT to solve the second subproblem in the ADMM scheme, respectively. Particularly, Fig.~\ref{fig:trajs_all_iters_in_all_solvers} (a) and (b) represent the trajectories solved by the SDP-ADMM and MIQP-ADMM, respectively. It's obvious that the number of iterations that the SDP-ADMM requires is much less than that of the MIQP-ADMM. Here, Fig.~\ref{fig:trajs_all_iters_in_all_solvers} (c) shows the trajectories with the use of one widely used interior point solver (IPOPT) to solve the second subproblem. It is straightforward that the IPOPT-ADMM requires much more iterations than both the SDP-ADMM and MIQP-ADMM.

\begin{figure}[t]
	\centering
	\includegraphics[width=1\linewidth]{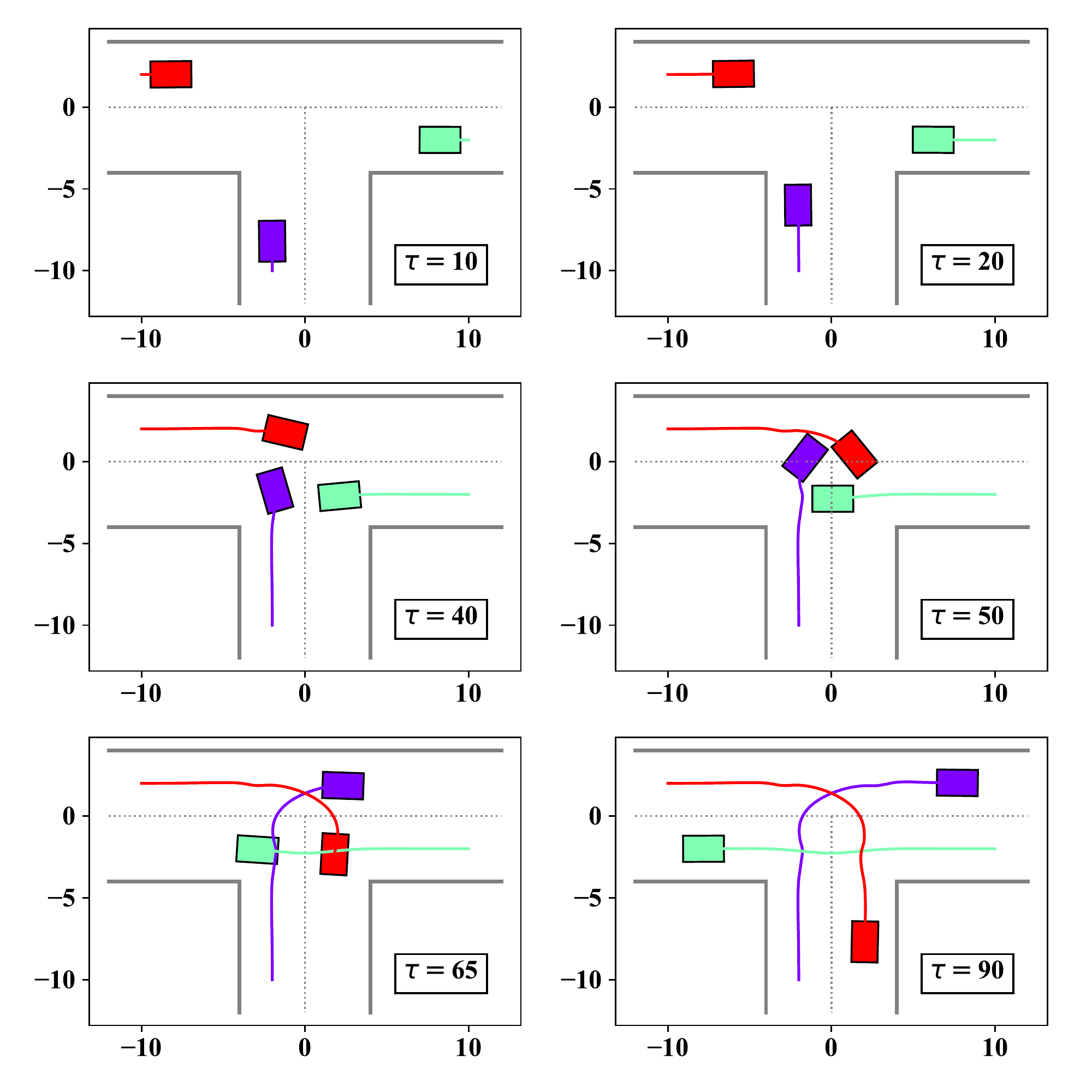}
	\caption{Trajectories at different time stamp $\tau$ for all vehicles in the last ADMM iteration .}
	\label{fig:trajs_diff_tau_in_T_SDP}
\end{figure}

Fig.~\ref{fig:trajs_diff_tau_in_T_SDP} is used to show how the three vehicles are controlled to reach the end points and how they avoid collision with each other with the use of the SDP-ADMM. The trajectories generated in the last ADMM iteration (which meets the predefined stopping criterion) are shown in Fig.~\ref{fig:trajs_diff_tau_in_T_SDP}. The six subfigures represent the current states of vehicles at the different time stamp $\tau$ in the final ADMM iteration. The curves with different colors denote the history trajectories from 0 to $\tau$. According to Fig.~\ref{fig:trajs_diff_tau_in_T_SDP}, all vehicles complete their driving task, meanwhile the collisions are avoided. Since the three vehicles' trajectories when using the MIQP-ADMM and IPOPT-ADMM are very similar to that of the SDP-ADMM, only the resulted driving process using the SDP-ADMM is demonstrated in this figure.  

\begin{figure}[t]
	\centering
	\includegraphics[width=0.8\linewidth]{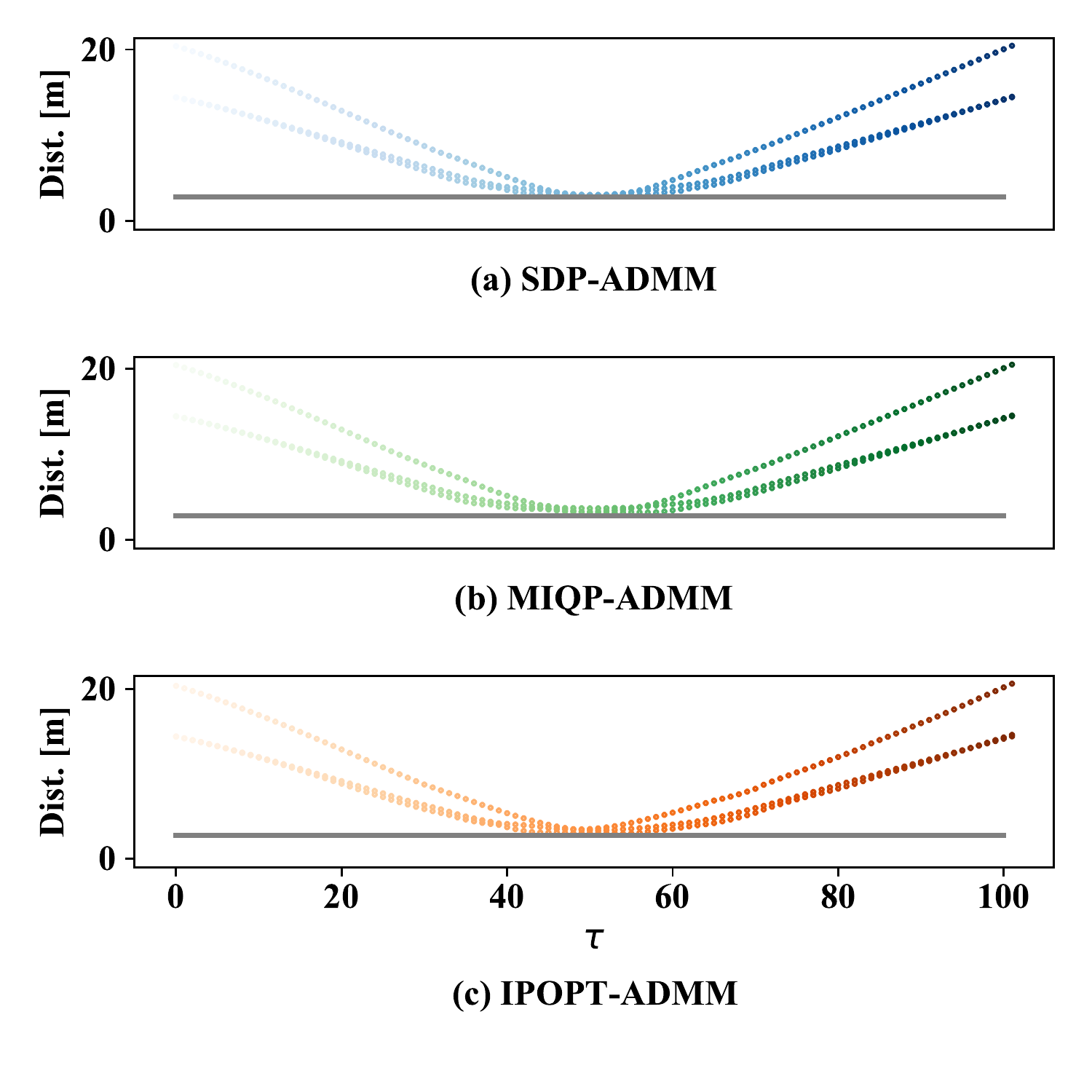}
	\caption{Distance among all vehicles in the last ADMM iteration.}
	\label{fig:Dist_all_solver_T}
\end{figure}

During the driving process, the safety distance among all vehicles should be maintained to avoid potential collisions. Here, Fig.~\ref{fig:Dist_all_solver_T} is used to show the distance among all vehicles in the last ADMM iteration under the three methods (SDP-ADMM, MIQP-ADMM, and IPOPT-ADMM) to solve the second subproblem. In this figure, the safety distance $d_{\text{safe}}=3$ m is represented by the gray solid line. Based on Fig.~\ref{fig:Dist_all_solver_T}, we can observe that the inter-distances for all vehicles are greater than the safety distance during the whole prediction horizon.

\subsubsection{Scenario 2 (Intersection)} 
\begin{figure}[t]
	\centering
	\includegraphics[width=1\linewidth]{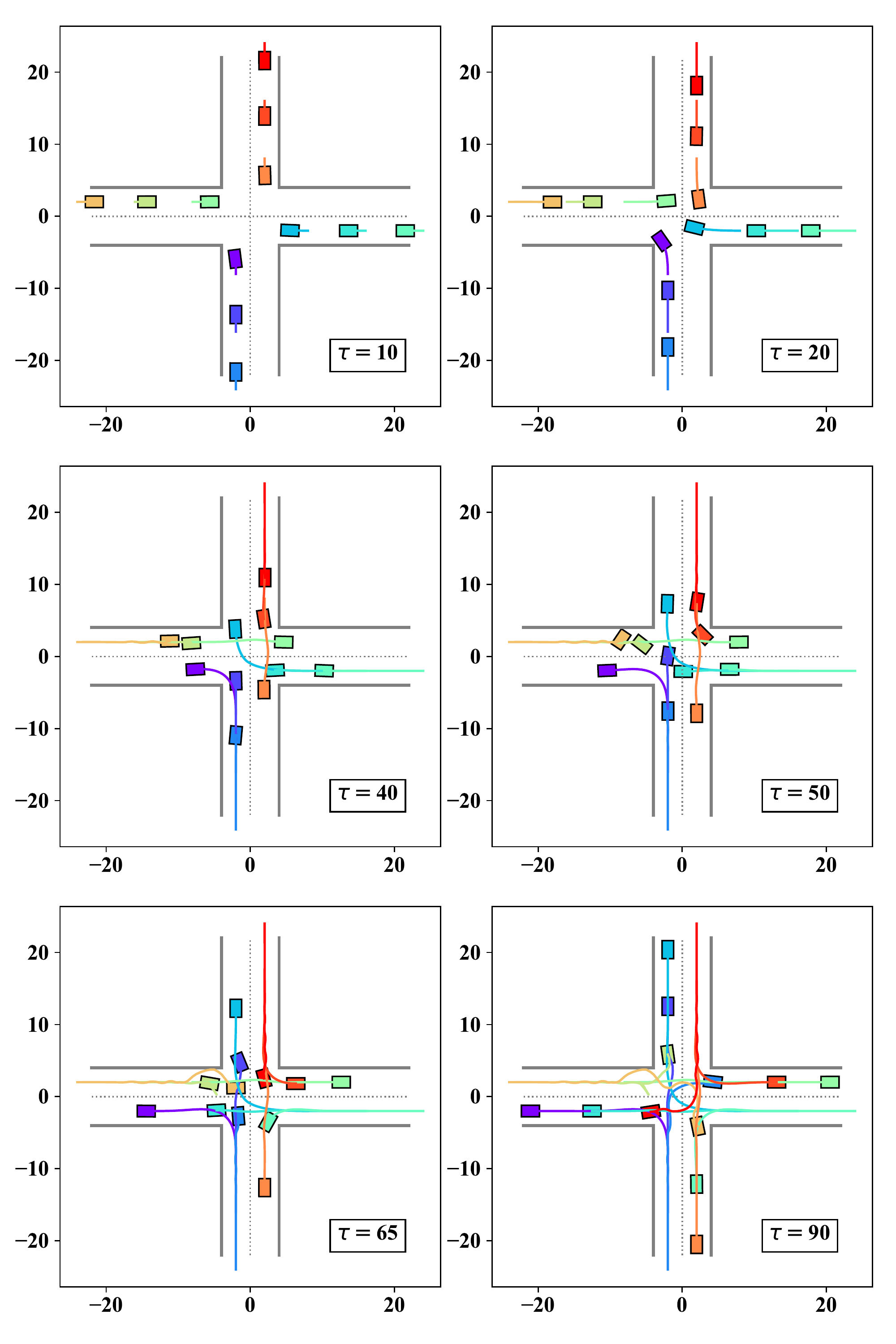}
	\caption{All trajectories at different time stamp $\tau$ for all vehicles in the last ADMM iteration with the use of SDP-ADMM.}
	\label{fig:trajs_diff_tau_in_X_SDP}
\end{figure}

In this scenario, there are 12 vehicles driving to pass through the intersection from 4 lanes. Note that vehicles are represented by using different colors. Similarly, there are 3 vehicles in one lane to carry out three driving behavior (turning right, turning left, and going straight). Fig.~\ref{fig:trajs_diff_tau_in_X_SDP} shows the driving process in  different time stamp $\tau$ for all vehicles, based on the last ADMM iteration when using the SDP-ADMM. It is easy to observe that all vehicles have successfully avoided each other by keeping a safe distance away. 

\begin{figure}[th]
	\centering
	\includegraphics[width=0.8\linewidth]{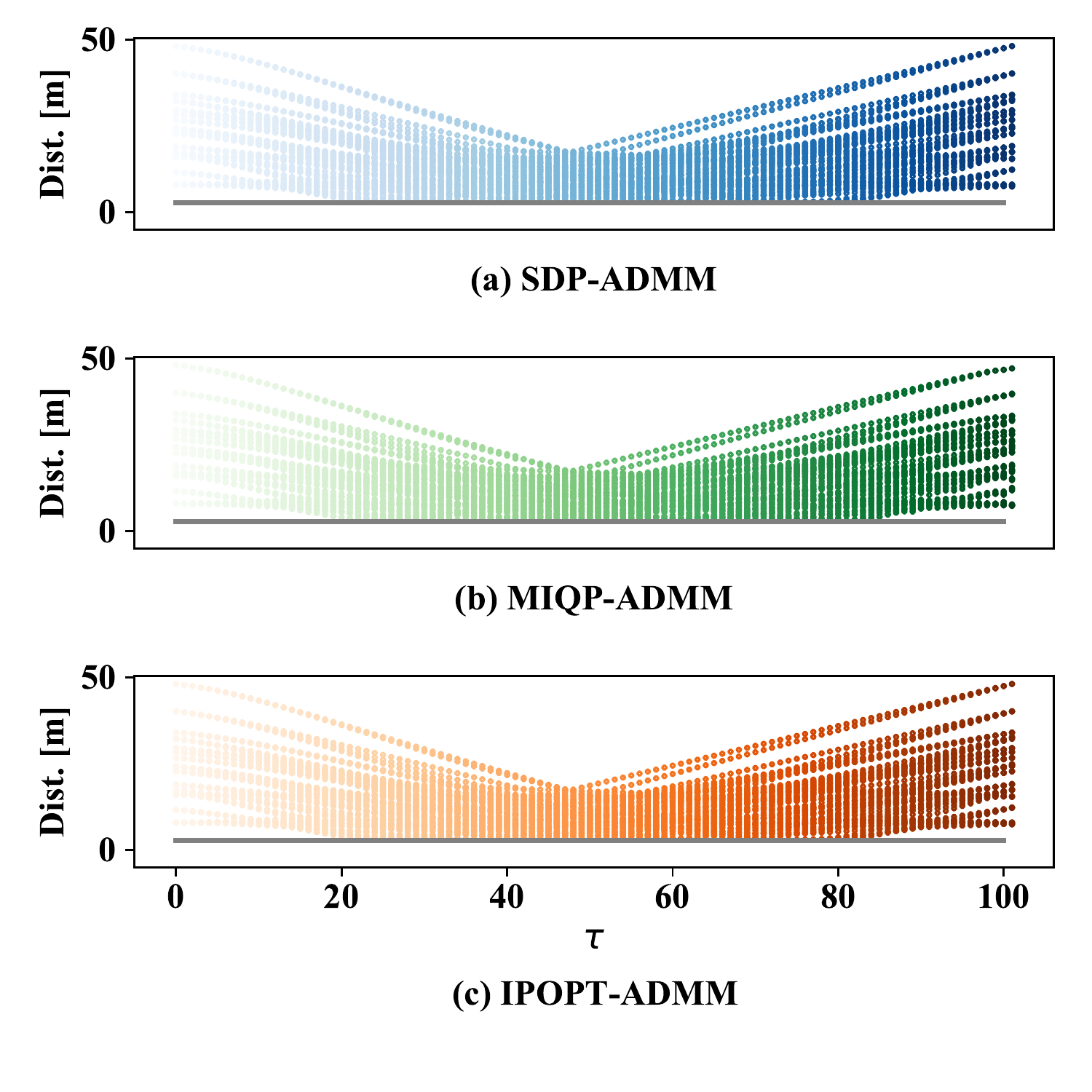}
	\caption{Distance among all vehicles in the last ADMM iteration.}
	\label{fig:Dist_all_solver_X}
\end{figure}

Similarly, Fig.~\ref{fig:Dist_all_solver_X} demonstrates the distance among all vehicles in the last ADMM iteration, under the three methods (SDP-ADMM, MIQP-ADMM, and IPOPT-ADMM) to solve the second subproblem. The gray solid line denotes the safety distance $d_{\text{safe}}=3$ m. Obviously, the inter-distances among all vehicles are greater than the safety distance during the whole prediction horizon.

\subsubsection{Comparison of Computational Time}

Table.~\ref{tab:iteration_number} shows the average iteration number of the SDP-ADMM, MIQP-ADMM, IPOPT-ADMM in the two driving scenarios for 20 trials. In this table, \#1 and \#2 represent the scenario 1 and scenario 2, respectively.  According to this table, the average iteration number of the SDP-ADMM is much less than that of the MIQP-ADMM and IPOPT-ADMM, which indicates the high computational efficiency of our proposed method. 
\begin{table}[thbp]
	\centering
	\caption{Average iteration number of the ADMM with use of three methods to solve the second subproblem in the two driving scenarios for 20 trials.}
	\begin{tabular}{|c|c|c|c|}
		\hline
		& SDP-ADMM & MIQP-ADMM & IPOPT-ADMM \\ \hline
		\#1 & 5   & 13   & 12    \\ \hline
		\#2 & 21  & 33   & 35    \\ \hline
	\end{tabular}
	\label{tab:iteration_number}
\end{table}

\begin{figure}[th]
	\centering
	\includegraphics[width=1\linewidth]{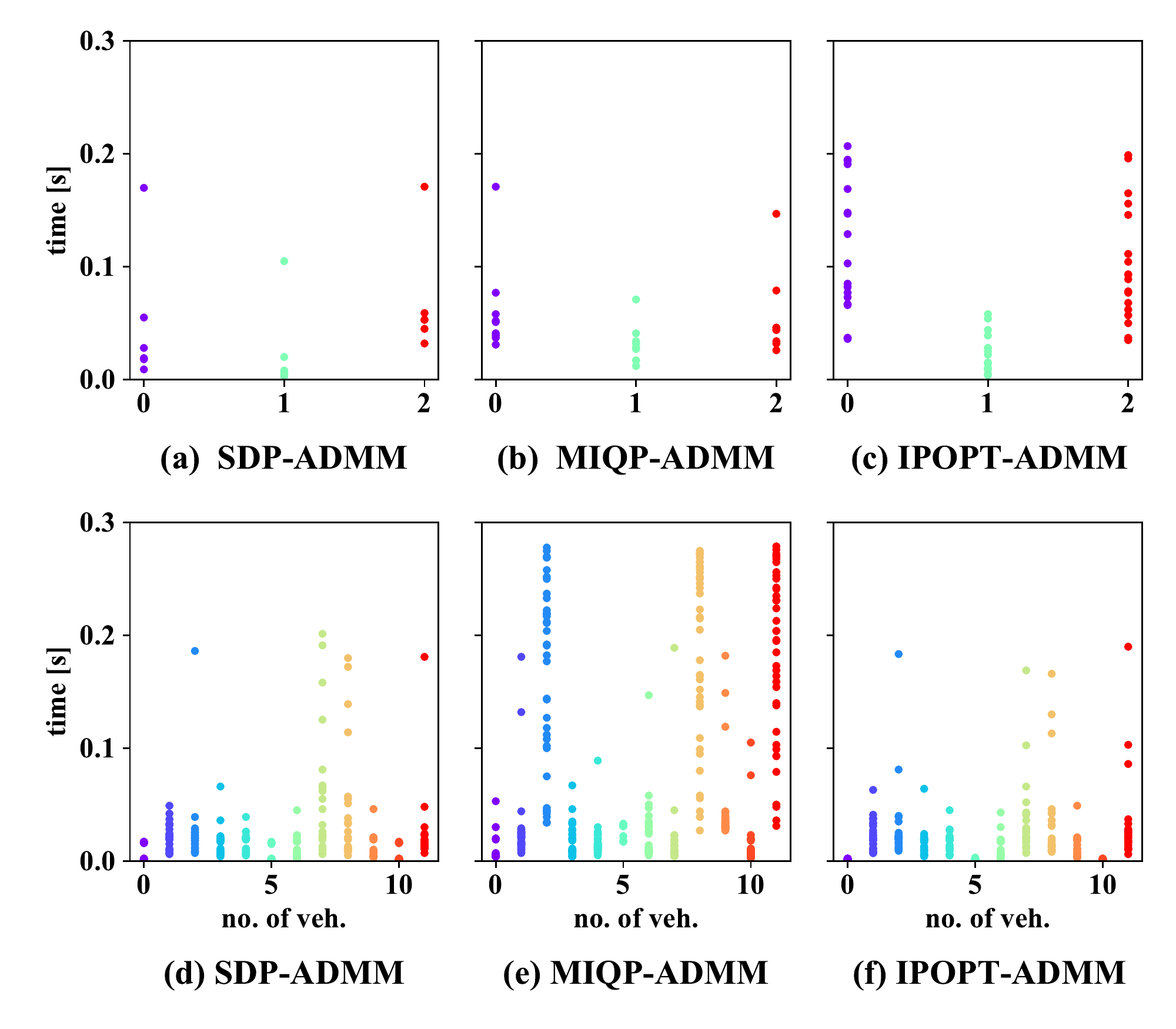}
	\caption{Computational time of solving the first subproblem in the ADMM for the three methods (the subfigures in the first row are the results in the scenario 1, and the subfigures in the second row shows the results in the scenario 2).}
	\label{fig:compare_first_step_time}
\end{figure}

\begin{figure}[th]
	\centering
	\includegraphics[width=1\linewidth]{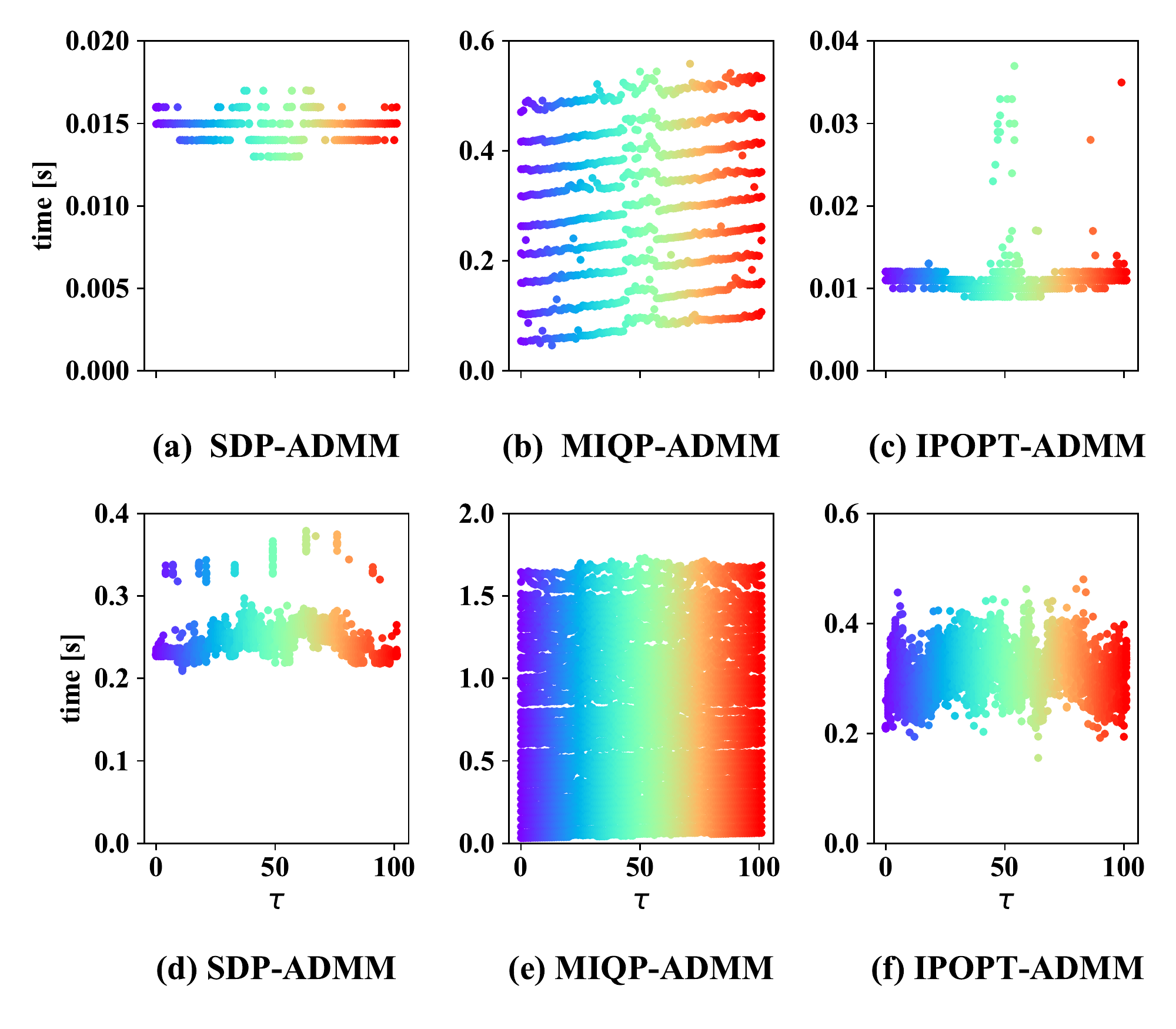}
	\caption{Computational time of solving the second subproblem in the ADMM for the three methods (the subfigures in the first row are the results in the scenario 1, and the subfigures in the second row shows the results in the scenario 2).}
	\label{fig:compare_second_step_time}
\end{figure}

\subsubsection{Comparison with Solvers}

Instead of using our proposed ADMM algorithm to compute in parallel, the original nonlinear and nonconvex problem~\eqref{eq:original_problem} can be solved by using a widely-used nonlinear programming solver, i.e., IPOPT. The approach which only use the IPOPT is called the pure-IPOPT. Note that here, the pure-IPOPT is compared with the whole ADMM-based solving approach. Here, the comparison method is to use the IPOPT to solve~\eqref{eq:original_problem}, instead of solving the second subproblem in the ADMM algorithm.

\begin{figure}[t]
	\centering
	\includegraphics[width=0.5\linewidth]{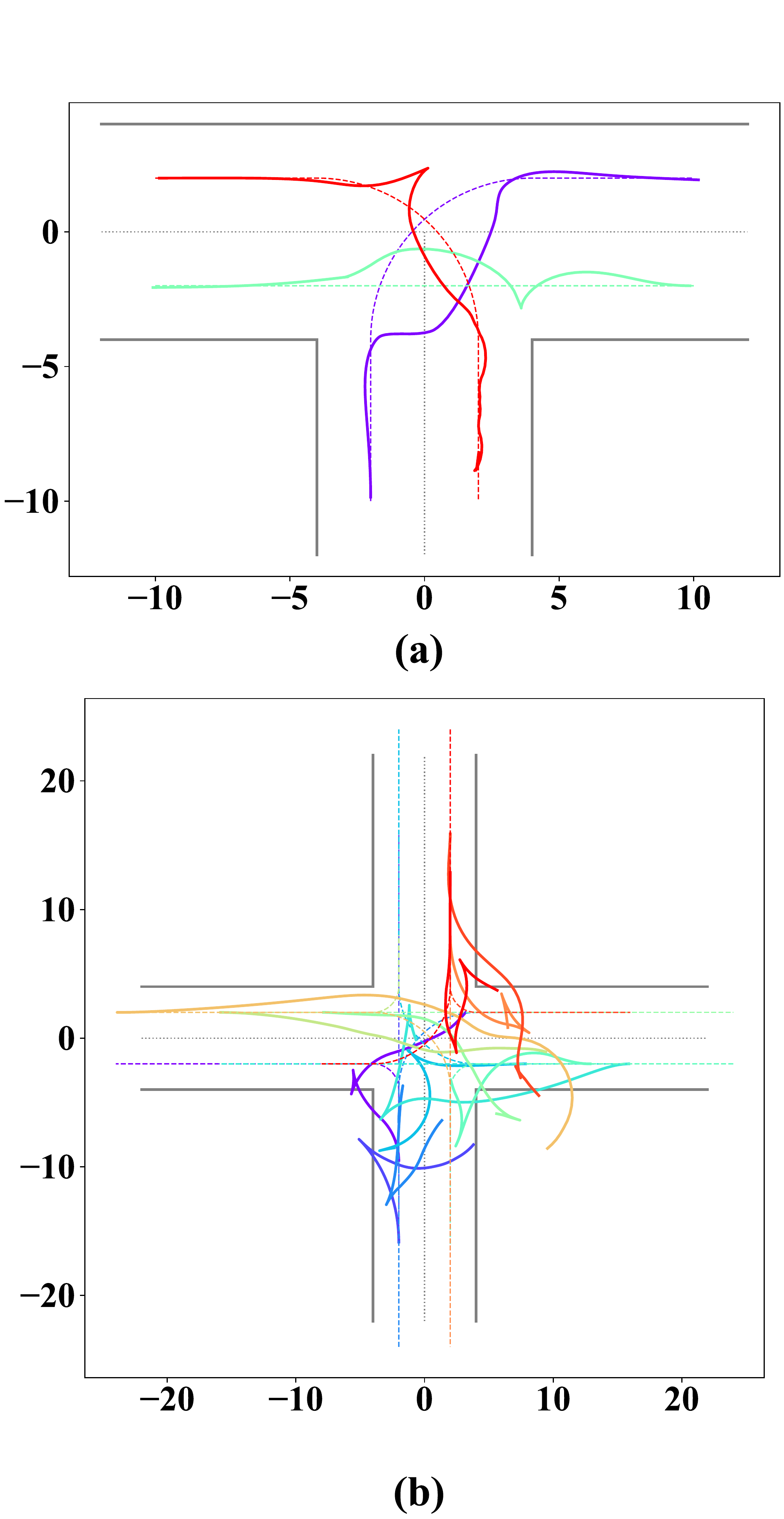}
	\caption{Trajectories for all vehicles by using pure-IPOPT to solve the original problem in the two scenarios.}
	\label{fig:trajs_centra_solver}
\end{figure}

The trajectories for all vehicles by using the IPOPT to solve the original problem in the two scenarios are shown in Fig.~\ref{fig:trajs_centra_solver} (a) and (b), respectively. In scenario 1 (three-way junction), the pure-IPOPT can successfully solve this problem, but it results in a low-quality solution with higher cost, compared with all of the three ADMM-based approaches. In scenario 2 of the intersection, the solution of the IPOPT solver has been trapped into a local minimum and cannot achieve the driving task successfully. On the country, our proposed approach can find a (sub-)optimal solution and finish the defined driving task successfully in both scenarios. Table~\ref{tab:comp_time} illustrates the comparison of the average computation time of the ADMM algorithm with use of three methods to solve the second subproblem and the pure-IPOPT which only use the IPOPT without the ADMM scheme in the two driving scenarios for 20 trials. From Table~\ref{tab:comp_time}, it is obvious that our proposed approach shows the best time efficiency, compared with the MIQP-ADMM, IPOPT-ADMM and pure-IPOPT. Note that the computation time of the pure-IPOPT in scenario 2 denotes the computational time of the unsuccessful trajectories, as shown in Fig.~\ref{fig:trajs_centra_solver} (b).

\begin{table}[thbp]
	\centering
	\caption{Comparison of the average computation time of the four methods (SDP-ADMM, MIQP-ADMM,IPOPT-ADMM, and pure-IPOPT) to solve the second subproblem in the two driving scenarios for 20 trials.}
	\begin{tabular}{|c|c|c|c|c|}
		\hline
		& SDP-ADMM & MIQP-ADMM & IPOPT-ADMM & pure-IPOPT   \\ \hline
		\#1 & 0.315   & 4.131   & 4.243 &  86.735  \\ \hline
		\#2 & 4.143  & 35.462   & 13.476  &  3078.512 \\ \hline
	\end{tabular}
	\label{tab:comp_time}
\end{table}

Our proposed approach (SDP-ADMM) also shows its effectiveness when solving such optimization problems, compared with the approach that only uses the SDP (which cannot address the nonlinear and nonconvex optimization problem~\eqref{eq:original_problem}). Certainly, we can use the SDR to relax the nonconvex constraints, but the nonlinear constraints, i.e., the dynamics constraints, cannot be handled. Besides, the dimension of the second subproblem~\eqref{eq:sub_problem2_4} in our proposed approach is much smaller than the original problem, which is contributed from the ADMM scheme by separating the original problem into two manageable subproblems and computing these subproblems in a parallel manner. A similar reason happens in the situation where the IPOPT-ADMM is much faster than the pure-IPOPT. Therefore, our proposed ADMM-based approach can achieve real-time computation due to the parallel computation and effective separation of the original optimization problem.

\section{Discussion and Conclusion}
\label{section:discussion_and_conclusion}
\subsection{Discussion}
Based on \cite{luo2010nonconvex}, the time complexity to solve the SDP in the worst cases is 
$\mathcal{O} (\max\{m, n\}^4 n^{\frac{1}{2}} \log(\frac{1}{\epsilon}))$, where $\epsilon\in \mathbb R_+$ denotes the numerical solution accuracy, $n$ is the dimension of decision variables and $m$ is the number of constraints. Note that the assumption of sparsity or specific structure in matrices, which can be used to improve the computation time by some solving tricks, is not considered in this time complexity. Thus, the SDP is a computationally efficient approximation method for the nonconvex QCQP, because the time complexity is polynomial time with the problem size $n$ and the number of constraints $m$.

As for the MIQP, the branch-and-cut algorithm is applied in most solvers. For the branch-and-cut algorithm, generally, all feasible solution sets are repeatedly divided into smaller and smaller subsets, which is called branches; and a target lower bound (for the minimum problem) is calculated for the solution set in each subset, which is called delimitation; After sub-branch, any subset whose limit exceeds the target value of the known feasible solution set will not be considered further, which is called pruning. As we all know, it is NP-complete; and thus, it's rather hard to provide the time complexity of the MIQP. 

\subsection{Conclusion}
This paper investigates the cooperative planning and control problem for multiple CAVs in autonomous driving. Here, a nonlinear nonconvex constrained optimization problem is suitably formulated, considering the nonlinear dynamics of the vehicle model and various coupling constraints (regarding the communications) among all CAVs. Next, we propose an ADMM-based approach to split the optimization problem into several small-scale subproblems, and these sub-problems can be efficiently solved in a parallel manner. Here, the nonlinearity of the system dynamics can be addressed efficiently by using the DDP algorithm, and the SDR approximates the nonconvexity of the coupling constraints with small dimensions, which can also be solved in a very short time. As a result, real-time computation and implementation can be realized through our proposed approach. Two complex driving scenarios in autonomous driving are used to validate the effectiveness and computational efficiency of our proposed approach in cooperative planning and control for multiple CAVs. 

\bibliographystyle{IEEEtran}
\bibliography{IEEEabrv,Reference}

\end{document}